\documentclass[
reprint,
superscriptaddress,
 amsmath,amssymb,
 aps,
prb,
]{revtex4-2}

\usepackage{dcolumn}
\usepackage{graphicx}
\usepackage{hyperref}
\usepackage{color}
\usepackage{tikz}
\usepackage[compat=1.1.0]{tikz-feynhand}
\usepackage{ascmac,amsmath,amsfonts,amsthm,amssymb}
\usepackage{url}
\usepackage{siunitx}
\usepackage[whole]{bxcjkjatype}

\usepackage{bm}

\newcommand{\n}{\nonumber \\}
\renewcommand{\t}[1]{\mathrm{#1}}

\def\ket#1{|#1\rangle }
\def\bra#1{\langle #1 |}
\def\braket#1{\langle #1 \rangle}
\def\n{\nonumber \\ }

\begin{document}

\preprint{APS/123-QED}

\title{Direct current generation by dielectric loss in ferroelectrics}

\author{Takahiro Morimoto}
\affiliation{Department of Applied Physics, The University of Tokyo, Hongo, Tokyo, 113-8656, Japan}

\author{Naoto Nagaosa}
\affiliation{RIKEN Center for Emergent Matter Science (CEMS), Wako, Saitama 351-0198, Japan}
\affiliation{Fundamental Quantum Science Program, TRIP Headquarters, RIKEN, Wako 351-0198, Japan}

\begin{abstract}
We study direct current (DC) generation induced by microwave irradiation to ferroelectric materials. The DC generation originates from microwave absorption called dielectric loss due to the delay of dielectric response. Such current generation can be formulated as the low-frequency limit of the phonon shift current which arises from an increase of electric polarization accompanying photoexcitation of phonons due to the electron-phonon coupling. 
To study the DC generation by the dielectric loss, we apply the diagrammatic treatment of nonlinear optical responses to photoexcitations of phonons and derive the general formula for phonon shift current. We then study the DC generation in the low-frequency region and find that the current scales as $\propto \omega^2$ for the linearly polarized light and time reversal symmetric systems. We estimate the order of magnitude of the DC generation by dielectric loss, indicating its feasibility for experimental detection in the GHz region.
\end{abstract}

\maketitle

\section{Introduction}\label{sec : Introduction}

\begin{figure}[t]
    \centering
    \includegraphics[width = \linewidth]{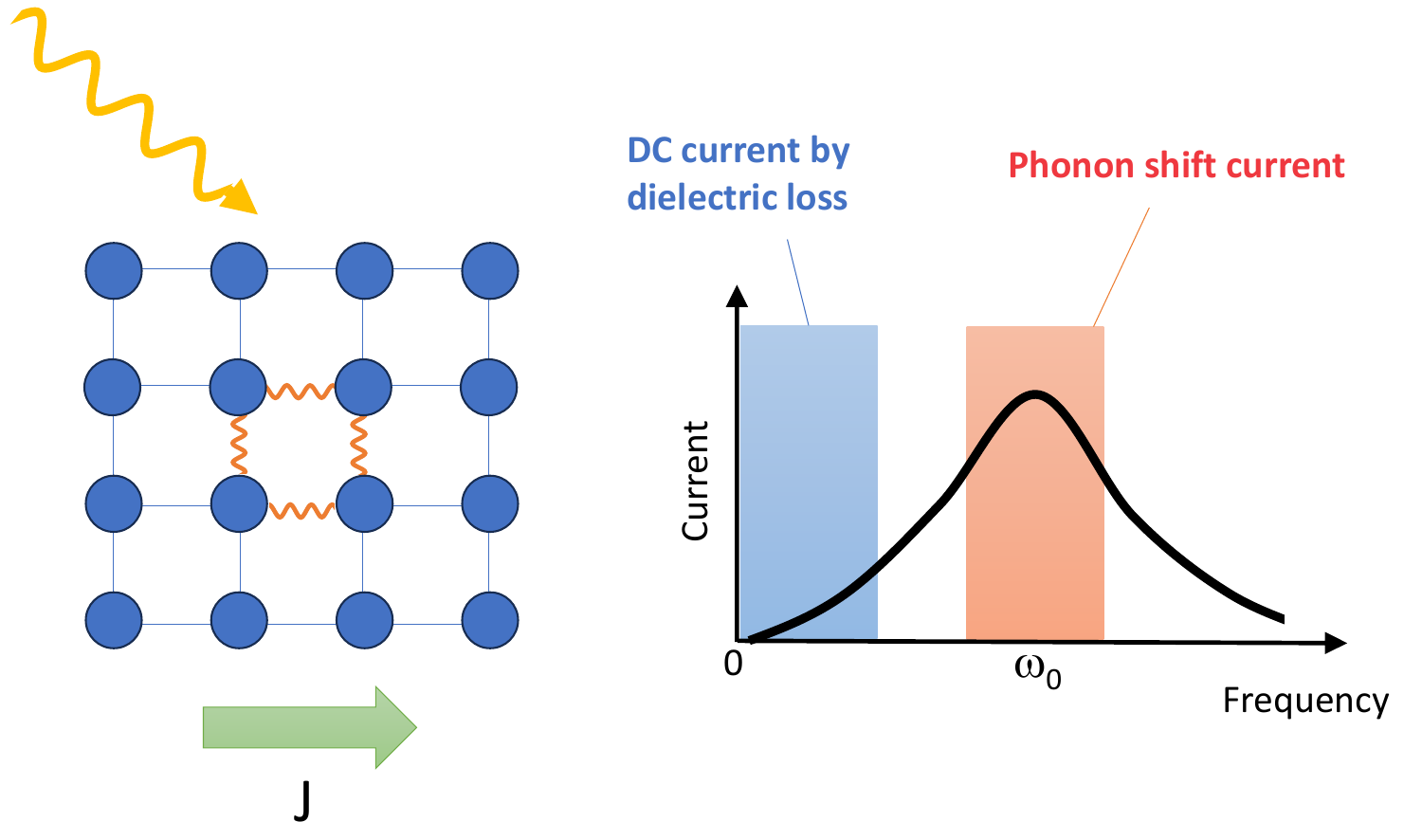}
    \caption{
    Schematic picture of direct current (DC) generation by dielectric loss. 
    (Left) Photoexcitation of phonons produces DC in inversion broken materials called phonon shift current, originating from an increase of electric polarization accompanying phonons due to electron-phonon coupling.
    (Right) In the low frequency region, off resonant phonon excitation leads to dielectric loss in ferroelectrics. According to the shift current mechanism, such dielectric loss also produces DC in the microwave regime.
    }
    \label{fig: schematics}
\end{figure}

Ferroelectricity is one of the most fundamental phenomena in solids, 
where the spontaneous electric polarization is induced by the 
atomic displacements breaking the spatial inversion symmetry \cite{Lines2001}.  
The simple picture is the separation between the positive and negative 
charges which results in the electric dipole and the electric polarization $P$.
The modern theory of electronic contribution to the 
polarization is formulated in terms of the quantum Berry phase of Bloch 
wave functions \cite{Resta,Kingsmith93}. 
For this formulation, the polarization current $j_P$ 
associated with the adiabatic change of the atomic positions
is the central quantity which is related to the Berry curvature of Bloch electrons.
The polarization is defined as the 
integral of $j_P$, resulting in Berry phase. Therefore, the ferroelectricity is directly 
relevant to the geometric nature of the electronic states in solids.
In recent years, it has been recognized that the Berry phase is also relevant to the 
second-order nonlinear optical response to produce the direct current (DC) in 
noncentrosymmetric materials, i.e., shift current \cite{Baltz,Sipe,Young-Rappe,Morimoto-Nagaosa16,Cook17,dai2022recent,Morimoto-JPSJ23}. 
Shift current is induced by the shift of the electronic wave-packet upon interband transition under light irradiation in noncentrosymmetric crystals. 
Namely, the photoexcited electron-hole pair has nonvanishing electric polarization due to the shift, and a constant increase of the electric polarization in time due to the photoexcitations of electron-hole pairs leads to the DC generation in the steady state.
Such shift of electrons can be
described by the Berry connection that quantifies the intracell coordinate of Bloch electrons. 
In this sense, shift current is of a geometric origin and is similar to the
polarization current.
In particular, it does not originate from the 
transport of the photo-induced carriers. 
That said, shift current is still in sharp contrast to the polarization current, in that shift current can be DC whereas the polarization current is always alternating current (AC) under alternating electric fields of electromagnetic waves
due to the reversal of the polarization.

Since the shift current does not require the transport of the
photo-induced carriers, it may not require the photocarriers themselves, i.e., any elementary excitation that accompanies nonzero electric polarization may induce DC generation once such excitation can be photoexcited.
This possibility has been explored in the exciton-induced shift current
both experimentally and theoretically \cite{Morimoto-exciton16,Sotome21,Chan21,Akamatsu21}. 
The photoexcitation of excitons
without the external bias voltage induces the DC shift current,
where no free charge carriers (i.e., free electrons and holes)
exist.
In this case, the photocurrent originates from finite electric polarization of excitons where the wave packets of bound electrons and holes are shifted.
Even more surprising is the discovery of the shift current
in ferroelectric material BaTiO$_3$ by the phonon excitation 
at THz region where no electronic excitations take place \cite{Okamura22}.
The DC emerges by photoexcitation of the phonon which is hybridized with the
virtual electron-hole pairs through the electron-phonon interaction.
In ferroelectric materials, such hybridization endows phonons with electric polarization, leading to DC generation with their photoexcitation.
Therefore, it is an intriguing issue how the photon energy can be reduced 
for the shift current generation in insulating materials.

In ferroelectrics, its AC response is described by the
dielectric function $\varepsilon(\omega) = \varepsilon'(\omega) 
+ i \varepsilon''(\omega)$. The imaginary part $\varepsilon''(\omega)$ corresponds to the 
energy dissipation which is proportional to $\omega$ in the low
frequency limit. 
This low energy response arises from the delay of the response, i.e., 
the phase shift, and is extensively studied in the microwave regime, called 
dielectric loss  \cite{Gurevich91,Sparks82,Subbaswamy86,Aupi04}. 

In the present paper, we study the relation between
the phonon-induced shift current and the dielectric loss in the low
frequency limit,
exploring the possible DC generation
associated with the dielectric loss (Fig.~\ref{fig: schematics}). 
To this end, we consider a system of Bloch electrons coupled to phonon excitations as summarized in Sec.~\ref{sec: setup}.
In Sec.~\ref{sec: derivation}, using a diagrammatic technique, we derive a general expression for the DC generation that arises from the second order nonlinear response in the external electric field and involves a phonon excitation as an intermediate process.   
This formulation gives a unified view for the phonon shift current and the DC generation by dielectric loss; the phonon shift current corresponds to the DC generation in the resonant regime ($\omega \simeq \omega_0$ with the phonon frequency $\omega_0$), and the DC generation by the dielectric loss corresponds to the low frequency and off-resonant regime ($\omega \ll \omega_0$),
as explained in Sec.~\ref{sec: phonon shift current} and Sec.~\ref{sec: dielectric loss}, respectively.
We demonstrate the DC generation in a representative 1D model of ferroelectrics, Rice-Mele model, with a coupling to phonon excitations in Sec.~\ref{sec: Rice Mele}, and give an order estimation of the DC generation in the low frequency ($\sim$ GHz) region in Sec.~\ref{sec: discussion}, revealing such DC generation is feasible for experimental detection in ferroelectric materials.

\section{Setup \label{sec: setup}}
In this section, we present our setup for DC generation, where Bloch electrons in solids are coupled to phonons and subjected to the external electric fields. 

We consider the Bloch electrons in solids described by the Hamiltonian,
\begin{align}
H_\t{el} &= \sum_{k,a}\epsilon_a(k) c_{k,a}^\dagger c_{k,a}, 
\end{align}
with the energy dispersion $\epsilon_a(k)$ and the annihilation operator of the Bloch electron $c_{k,a}$ for the band $a$ and the momentum $k$.
The coupling to the external electric field is introduced by the minimal coupling as  \cite{Parker19}
\begin{align}
H_\t{el-A} &= e\sum_{k,a,b,\alpha}
A^\alpha v_{ab}^\alpha(k) c_{k,a}^\dagger c_{k,b} \n
&+ e^2 \sum_{k,a,b,\alpha,\beta} A^\alpha A^\beta (\partial_{k_\alpha} v^\beta)_{ab}(k) c_{k,a}^\dagger c_{k,b} +O(A^3), 
\end{align}
where we set $\hbar=1$ for simplicity.
Here, $A^\alpha$ is the vector potential of the electric field along the $\alpha$ direction and we only consider contributions up to $O(A^2)$ as we focus on the second order effect.
The velocity operator $v^\alpha$ along the $\alpha$ direction is given by $v^\alpha=\partial_{k_\alpha} H$ when $H$ is represented with $k$ independent basis wave functions.
Similarly, the diamagnetic current operator $\partial_{k_\alpha} v^\beta$ is given by 
$\partial_{k_\alpha} v^\beta =\partial_{k_\alpha}\partial_{k_\beta} H$ 
in the same representation \cite{Morimoto-JPSJ23}.
In general, one can formulate those current operators as the covariant derivative of the Hamiltonian $v^\alpha=\mathcal{D}^\alpha H$ and that of the velocity operator $\partial_{k_\alpha} H = \mathcal{D}^\alpha H$. Here the covariant derivative $\mathcal{D}^\alpha$ of an operator $O$ is defined with the matrix elements $(\mathcal{D}^\alpha O)_{ab}= \partial_{k_\alpha} O_{ab} - i [\mathcal{A}^\alpha,O]_{ab}$ with the Berry connection $\mathcal{A}^\alpha=i\braket{u_a|\partial_{k_\alpha} u_b}$, where $\ket{u_a}$ is the periodic part of the Bloch wave function of the band $a$,
and the operators with subscripts denote their matrix elements with the Bloch states $O_{ab}=\braket{u_a|O|u_b}$.
The interband matrix element of the Berry connection is written with a matrix element of the velocity operator as
$\mathcal{A}^\alpha_{ab}=-iv_{ab}^\alpha/\epsilon_{ab}$ with $\epsilon_{ab}=\epsilon_a - \epsilon_b$.

The electron-phonon coupling is given by \cite{Mahan}
\begin{align}
H_\t{el-ph} &= \frac{1}{\sqrt{V}} \sum_{k,q,a,b}g_{ab}(k+q,k) c_{k+q,a}^\dagger c_{k,b} (b_q + b_{-q}^\dagger ), 
\end{align}
where $V$ is the total volume of the system, 
$g_{ab}(k+q,k)$ is the matrix element of the electron phonon coupling in which the electron momentum is modified from $k$ to $k+q$ with the phonon momentum $q$,
 $c_{k,a}$ is the annihilation operator of electrons with the momentum $k$ and the band index $a$, and $b_q$ is the annihilation operator of phonons with the momentum $q$.
The summation runs discrete momenta given by $k=2\pi n/L$ with the linear dimension $L$ and an integer vector $n \in \mathbb{Z}^3$.
Since we consider photoexcitation of phonons, the momentum transfer $q$ is zero.
In addition, we further consider the modulation of the electron-phonon coupling $g(k)$ under the electromagnetic field $A^\alpha$, which is incorporated with a minimal substitution $g(k) \to g(k)+e\sum_\alpha A^\alpha \partial_{k_\alpha} g(k)$ (where we abbreviated $g_{ab}(k,k)$ by $g_{ab}(k)$). 
This modulation term $\propto \partial_{k_\alpha} g$ along with the diamagnetic current term in Eq. (2) plays an important role to obtain the correct behavior of the DC generation in the low frequency region avoiding an unphysical divergence.
Thus the relevant part of the electron-phonon coupling is given by
\begin{align}
H_\t{el-ph} &= \frac{1}{\sqrt{V}} \sum_{k,a,b} [g_{ab}(k) + e\sum_\alpha A^\alpha (\partial_{k_\alpha} g)_{ab}(k)]
\n
&\qquad\qquad \times c_{k,a}^\dagger c_{k,b} (b_0 + b_{0}^\dagger).
\label{eq: Helph q=0}
\end{align}
We note that one may adopt the different notation for the electron-phonon coupling given by
\begin{align}
H_\t{el-ph} &= \frac{1}{\sqrt{N}} \sum_{k,q,a,b}\tilde g_{ab}(k+q,k) c_{k+q,a}^\dagger c_{k,b} (b_q + b_{-q}^\dagger ), 
\end{align}
where $N$ is the number of unit cells within the system (i.e., $N=V/V_c$ with the unit cell volume $V_c$).
In this notation, $\tilde g$ has the dimension of energy and is related to the former notation via $g(k,k') = \sqrt{V_c} \tilde g(k,k')$.

\section{DC generation with phonons \label{sec: derivation}}
In this section, we present the diagrammatic formulation of DC generation by phonon excitations \cite{Okamura22}.
When the photon energy $\hbar \omega$ is resonant with the phonon energy $\hbar \omega_0$, the real excitation of phonons takes place and leads to DC generation called phonon shift current. 
When the photon energy $\hbar \omega$ is off resonant with the phonon energy $\hbar \omega_0$, e.g., $\hbar \omega \ll \hbar \omega_0$, the energy dissipation still takes place due to finite energy broadening of the phonon spectrum and results in nonzero DC generation. 

\begin{figure}
\begin{center}
\includegraphics[width=\linewidth]{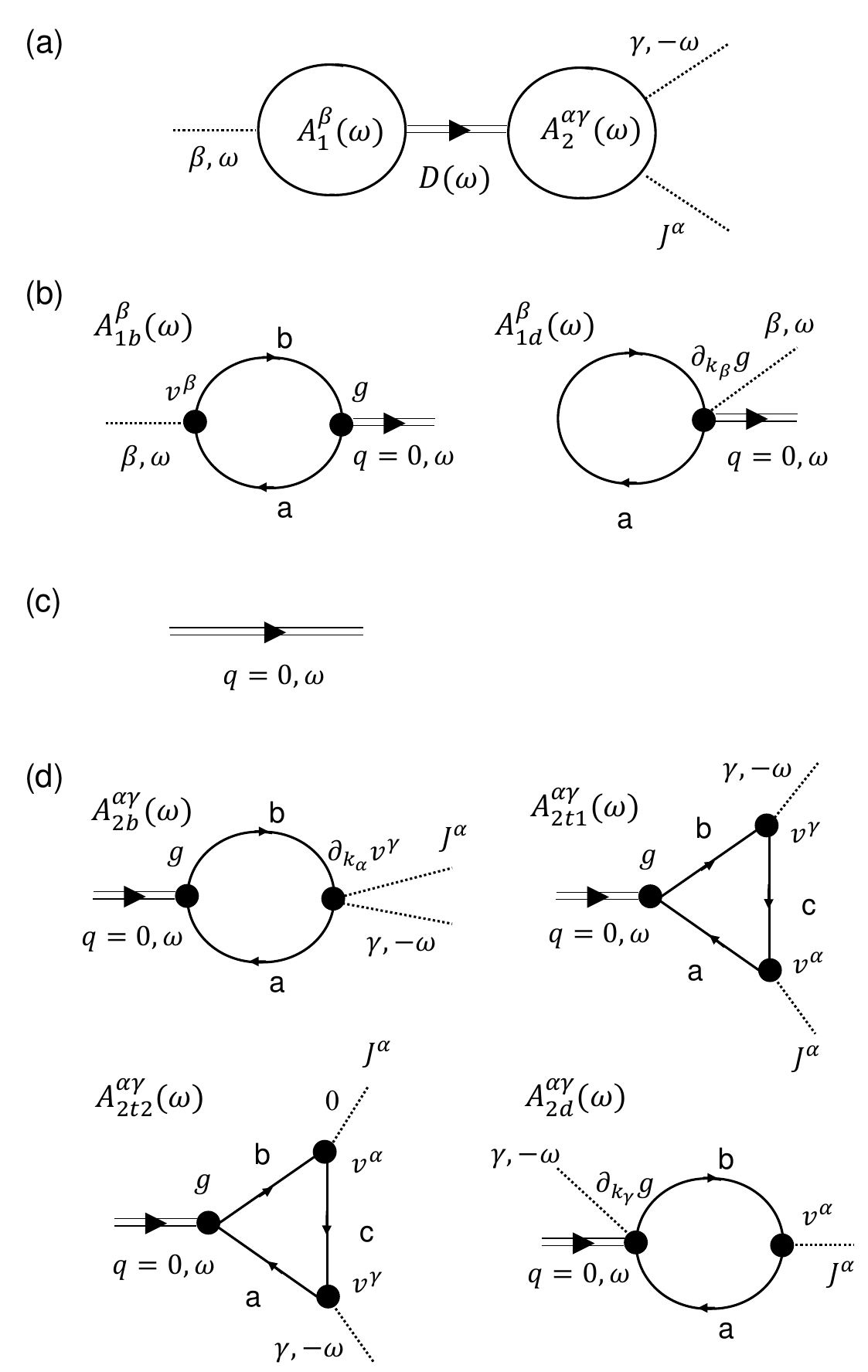}
\caption{\label{fig: diagrams}
Diagrams for direct current (DC) generation induced by phonon excitations.
(a) The second order nonlinear process responsible for DC generation with phonon excitations.
(b) Diagrams $A_1^\beta(\omega)$ with incoming photon with the frequency $\omega$ and outgoing phonon.
(c) Phonon propagator $D(\omega)$.
(d) Diagrams $A_2^{\alpha\gamma}(\omega)$ with incoming phonon, incoming photon with the frequency $-\omega$ and the current vertex.
}
\end{center}
\end{figure}

\subsection{Diagrammatic derivation of the nonlinear conductivity}

We consider DC generation at the second order of the external electric field, which is described by the nonlinear conductivity $\sigma_{\alpha \beta \gamma}(\omega)$ as
\begin{align}
J_\t{dc}^\alpha = \sigma_{\alpha \beta \gamma}(\omega) E^\beta(\omega) E^\gamma(-\omega),
\label{eq: J shift}
\end{align}
where $J_\t{dc}^\alpha$ is DC along the $\alpha$ direction and $E^\alpha(\omega)$ is the electric field along the $\alpha$ direction of the frequency $\omega$.
The nonlinear conductivity $\sigma_{\alpha \beta \gamma}(\omega)$ can be obtained using the standard diagrammatic technique as \cite{Mahan,Parker19}
\begin{align}
\sigma_{\alpha \beta \gamma}(\omega) &=
\frac{e^3}{\omega^2} \Pi_{\alpha \beta \gamma}(\omega),
\label{eq: sigma abc}
\end{align}
with the response function $\Pi_{\alpha \beta \gamma}$.
If we focus on the current generation which arises from the phonon excitations and is in the lowest order in the electron-phonon coupling $g$,
the response function is given by 
\begin{align}
\Pi_{\alpha \beta \gamma}(\omega)
&=
A^\beta_1(\omega) D(\omega) A^{\alpha\gamma}_2(\omega) +
A^\gamma_1(-\omega) D(-\omega) A^{\alpha\beta}_2(-\omega),
\end{align}
where the diagrammatic representation for the first term is illustrated in Fig. \ref{fig: diagrams}(a).
Here, 
$A_1^\beta(\omega)$ represents contributions from diagrams with incoming photon with the frequency $\omega$ and outgoing phonon (Fig. \ref{fig: diagrams}(b)).
The phonon propagator $D(\omega)$ is given by
\begin{align}
D(\omega) &= \frac{2\omega_0}{\omega^2 - \omega_0^2 + 2i\gamma \omega} \n
&\simeq \frac{1}{\omega - \omega_0 + i\gamma} - \frac{1}{\omega + \omega_0 +i\gamma},
\end{align}
with the phonon frequency $\omega_0$ and the energy broadening for the phonon $\gamma$ (Fig. \ref{fig: diagrams}(c)) \cite{Scott74}.
$A_2^{\alpha\gamma}(\omega)$ represents contributions from the diagrams with incoming phonon, incoming photon with the frequency $-\omega$ and the current vertex (Fig. \ref{fig: diagrams}(d)).

To compute contributions from the diagrams, we adopt the imaginary time formalism for Green's function. For simplicity, we focus on the zero temperature case, where the Green's function in the Matsubara frequency representation is given by
\begin{align}
    G(k,i\omega)&=\sum_a \frac{\ket{u_a} \bra{u_a}}{i\omega- \epsilon}.
\end{align}
We consider two incoming photons with Matsubara frequencies $i\Omega_1$ and $i\Omega_2$ that are analytically continued as 
\begin{align}
    i\Omega_1 &\to \omega +i \gamma',
    &
    i\Omega_2 &\to -\omega +i \gamma',
\end{align}
with the photon frequency $\omega$ and the relaxation strength $\gamma'$. 
Below, we assume the relaxation strength $\gamma'$ is much smaller than the band separation of electrons, and we neglect $\gamma'$.
The first bubble diagram $A_{1b}^\beta$ is given by
\begin{align}
A^\beta_{1b}(i\Omega_1) &= \int \frac{d\omega}{2\pi} \int[dk] \t{Tr}[g G(k, i\omega + i \Omega_1) v^\beta G(k, i \omega)] \n
&= \int \frac{d\omega}{2\pi} \int[dk] \sum_{ab} g_{ab}\frac{1}{i\omega+i\Omega_1-\epsilon_a} v^\beta_{ba} \frac{1}{i\omega-\epsilon_a} \n
&=\int[dk] \sum_{ab} \frac{f_{ab} g_{ab} v^\beta_{ba}}{i\Omega_1-\epsilon_{ba}}.
\end{align}
Here we use the notation $\int [dk]\equiv \int d^dk/(2\pi)^d$, $\epsilon_a$ and $f_a$ are the energy and the Fermi distribution function for the state $a$, respectively.
We also use the notations: $f_{ab}=f_a-f_b$ and $\epsilon_{ab}=\epsilon_a - \epsilon_b$.
In addition, we have a contribution $A_{1d}^\beta$ arising from modulation of electron-phonon coupling $g$ in the presence of the electric field, which is described by the vertex $\partial_k g$ in a similar manner to the diamagnetic current (Fig.~\ref{fig: diagrams}(b)).
This contribution is important to derive correct behavior of the response function in the low frequency region, avoiding unphysical divergence that arises from $1/\omega^2$ factor in Eq.~\eqref{eq: sigma abc}.
Specifically, this contribution reads
\begin{align}
A^\beta_{1d}(i\Omega_1) &= \int \frac{d\omega}{2\pi} \int[dk] \t{Tr}[\partial_{k_\beta} g G(k, i \omega)] \n
&= -\int \frac{d\omega}{2\pi} \int[dk] \t{Tr}[g G(k, i\omega) v^\beta G(k, i \omega)] \n
&=-A^\beta_{1}(0).
\end{align}
Combining these two contributions and performing the analytic continuation $i\Omega_1 \to \omega$ and writing $\tilde A_1^\beta = A_1^\beta/\omega$, we obtain
\begin{align}
    \tilde A_1^\beta(\omega) 
    &= \frac{1}{\omega} (A^\beta_{1b}(\omega)+A^\beta_{1d}(\omega)) \n
    &=\int[dk] \sum_{ab} \frac{1}{\epsilon_{ba}}\frac{f_{ab} g_{ab} v^\beta_{ba}}{\omega-\epsilon_{ba}},
    \label{eq: A1}
\end{align}
by using the identity,
\begin{align}
    \frac{1}{\omega}\left(
    \frac{1}{\omega-x}-\frac{1}{-x}
    \right)
    &=\frac{1}{x(\omega-x)}.
\end{align}

The second bubble diagram $A^{\alpha\gamma}_2$ has three contributions: a bubble diagram made of diamagnetic current vertex $A_{2b}^{\alpha\gamma}$, a triangle diagram involving two (paramagnetic) current vertices $A_{2t}^{\alpha\gamma}$,
and a bubble diagram $A_{2d}^{\alpha\gamma}$ involving $\partial_k g$.
The diagram $A_{2b}^{\alpha\gamma}$ can be computed in a similar manner as $A_1^\beta$ and is given by
\begin{align}
A_{2b}^{\alpha\gamma}(i\Omega_1) &= \int \frac{d\omega}{2\pi} \int[dk] \t{Tr}[\partial_{k_\alpha} v^\gamma G(k, i\omega + i \Omega_1) g G(k, i \omega)] \n
&= \int[dk] \sum_{ab}  \frac{f_{ab} (\partial_{k_\alpha} v^\gamma)_{ab} g_{ba}}{i\Omega_1-\epsilon_{ba}}.
\end{align}
The diagram $A_{2t}^{\alpha\gamma}$ can be further separated into two pieces $A_{2t1}^{\alpha\gamma}$ and $A_{2t2}^{\alpha\gamma}$ according to the order of the other incoming photon and the current vertex.
The contributions $A_{2t1}^{\alpha\gamma}$ is given by
\begin{align}
A_{2t1}^{\alpha\gamma}(i\Omega_1) &=
\int \frac{d\omega}{2\pi} \int[dk] \t{Tr}[v^\alpha G(k, i\omega+ i \Omega_1+ i \Omega_2) \n
&\hspace{6em} \times 
v^\gamma G(k, i\omega + i \Omega_1) g G(k, i \omega)] \n
&=\int[dk] \sum_{a,b,c} v^\alpha_{ac} v^\gamma_{cb} g_{ba} I_{abc}(i\Omega_1, i\Omega_2),
\end{align}
with
\begin{align}
& I_{abc}(i\Omega_1,i\Omega_2) \n
&=\int \frac{d\omega}{2\pi} \frac{1}{(i\omega+i\Omega_1+i\Omega_2-\epsilon_c)(i\omega+i\Omega_1-\epsilon_b)(i\omega-\epsilon_a)} \n
&=\frac{1}{\epsilon_{ac}+i\Omega_1+i\Omega_2}\left(\frac{f_{ab}}{i\Omega_1-\epsilon_{ba}} - \frac{f_{cb}}{-i\Omega_2 -\epsilon_{bc}}\right).
\end{align}
Similarly, the contributions $A_{2t2}^{\alpha\gamma}$ is given by
\begin{align}
&A_{2t2}^{\alpha\gamma}(i\Omega_1) \n
&=
\int \frac{d\omega}{2\pi} \int[dk] \t{Tr}[v^\gamma G(k, i\omega - i\Omega_2) v^\alpha G(k, i\omega + i \Omega_1) g G(k, i \omega)] \n
&=\int[dk] \sum_{a,b,c} v^\gamma_{bc} v^\alpha_{ca} g_{ab} I_{bac}(i\Omega_1,-i\Omega_1-i\Omega_2) \n
&=\int[dk] \sum_{a,b,c} (v^\alpha_{ac} v^\gamma_{cb} g_{ba})^* I_{abc}(-i\Omega_1,-i\Omega_2),
\end{align}
where we 
changed the variable as $i\omega \to i\omega-i \Omega_1$ in the integral $I$ in the last line.
In addition, the contribution from $A_{2d}^{\alpha\gamma}$ that involves $\partial_k g$ reads
\begin{align}
    A_{2d}^{\alpha\gamma}(i\Omega_1) &= \int \frac{d\omega}{2\pi} \int[dk] \t{Tr}[v^\gamma G(k, i\omega) \partial_{k_\alpha} g G(k, i \omega)] \n
&=- \int \frac{d\omega}{2\pi} \int[dk] \t{Tr}[\partial_{k_\alpha} v^\gamma G(k, i\omega) g G(k, i \omega)
\n
&\hspace{2em}
+
v^\alpha G(k, i\omega) v^\gamma G(k, i\omega) g G(k, i \omega)
\n
&\hspace{2em}
+
v^\gamma G(k, i\omega) v^\alpha G(k, i\omega) g G(k, i \omega)
], \n
&= -(A_{2b}(0)+A_{2t1}(0)+ A_{2t2}(0)),
\end{align}
which correctly cancels the low frequency divergences of $A_{2b}, A_{2t1}$, and $A_{2t2}$.

\begin{widetext}
Thus after the analytic continuation of Matsubara frequencies and writing $A_2^{\alpha\gamma} = A_2^{\alpha\gamma}/\omega$, we obtain
\begin{align}
\tilde A_2^{\alpha\gamma}(\omega)
&= \frac{1}{\omega} (A_{2b}^{\alpha\gamma}(\omega)+A_{2t1}^{\alpha\gamma}(\omega)+ A_{2t2}^{\alpha\gamma}(\omega) + A_{2d}^{\alpha\gamma}(\omega)) \n
&= \int[dk] \sum_{ab}  \frac{f_{ab} (\partial_{k_\alpha} v^\gamma)_{ab} g_{ba}}{\epsilon_{ba}(\omega-\epsilon_{ba})} \n
&+ \int[dk] \sum_{abc} v^\alpha_{ac} v^\gamma_{cb} g_{ba} \frac{1}{\epsilon_{ac}}\left(\frac{f_{ab}}{\epsilon_{ba}(\omega-\epsilon_{ba})} - \frac{f_{cb}}{\epsilon_{bc}(\omega-\epsilon_{bc})}\right) \n
&+ \int[dk] \sum_{abc} (v^\alpha_{ac} v^\gamma_{cb} g_{ba})^* \frac{1}{\epsilon_{ac}}
\left(-\frac{f_{ab}}{\epsilon_{ba}(-\omega-\epsilon_{ba})} + \frac{f_{cb}}{\epsilon_{bc}(-\omega - \epsilon_{bc} )}\right)
.
\label{eq: A2}
\end{align}
\end{widetext}

The DC that can be extracted out of the system inevitably accompanies energy dissipation in the system. In the present case, such energy dissipation arises from the phonon excitation and is described by the imaginary part of the phonon propagator $\t{Im}[D(\omega)]$.
Thus, focusing on the contributions that involve $\t{Im}[D(\omega)]$,
we obtain the general expression for the DC generation with phonon exictations as
\begin{align}
    \sigma_{\alpha \beta \gamma}(\omega) 
    &=
    -e^3 C_{\alpha\beta\gamma}(\omega) \t{Im}[D(\omega)],
    \label{eq: sigma Im D}
\end{align}
with
\begin{align}
    C_{\alpha\beta\gamma}(\omega)
    &=
    \t{Im}\left[ \tilde A^\beta_1(\omega) 
    \tilde A^{\alpha\gamma}_2(\omega)
    -
    \tilde A^\gamma_1(-\omega) \tilde A^{\alpha\beta}_2(-\omega)
    \right]. 
    \label{eq: C}
\end{align}

\subsection{Role of time reversal symmetry}

Next let us consider the role of the time reversal symmetry $\mathcal T$ on the nonlinear conductivity $\sigma_{\alpha \beta \gamma}(\omega)$.
When the system preserves $\mathcal{T}$,
the energy dispersion is $k$-even as 
\begin{align}
    \epsilon_a(k)=\epsilon_a(-k),
\end{align}
and the matrix elements of the velocity operator $v$, the diamagnetic current $\partial_k v$ and the electron-phonon coupling $g$ satisfy the following relationships,\begin{align}
 v_{ab}(k) &= -v_{ba}(-k), \n
 (\partial_k v)_{ab}(k) &= (\partial_k v)_{ba}(-k), \n
 g_{ab}(k) &= g_{ba}(-k).
\end{align}
Note that the Bloch states $u_a(\pm k)$ are Kramers pairs for cases with spin orbit couplings.
With these symmetry properties, $\tilde A_1^\beta$ becomes purely imaginary and $\tilde A_2^{\alpha\gamma}$ becomes purely real.
Expanding $\tilde A_1^\beta$ and $\tilde A_2^{\alpha\gamma}$ with respect to $\omega$ gives
\begin{align}
    \tilde A_1^\beta 
    &=-i\int[dk] \sum_{ab} \t{Im}[g_{ab} v^\beta_{ba}] \frac{f_{ab}}{\epsilon_{ba}^2} \n
    & -i \omega \int[dk] \sum_{ab} \t{Im}[g_{ab} v^\beta_{ba}] \frac{f_{ab}}{\epsilon_{ba}^3}
    +O(\omega^2),
    \label{eq: A1 TRS}
\end{align}
and
\begin{align}
\tilde A_2^{\alpha\gamma} 
&= -\int[dk] \sum_{ab}  
\t{Re}[(\partial_{k_\alpha} v^\gamma)_{ab} g_{ba}]
\frac{f_{ab} }{\epsilon_{ba}^2} \n
& -\omega \int[dk] \sum_{ab}  
\t{Re}[(\partial_{k_\alpha} v^\gamma)_{ab} g_{ba}]
\frac{f_{ab} }{\epsilon_{ba}^3} \n
&-2\omega \int[dk] \sum_{abc} 
\t{Re}[v^\alpha_{ac} v^\gamma_{cb} g_{ba}]
\frac{1}{\epsilon_{ac}}
\left(\frac{f_{ab}}{ \epsilon_{ba}^3} 
- 
\frac{f_{cb}}{\epsilon_{bc}^3}
\right) 
\n
&+O(\omega^2)
.
\label{eq: A2 TRS}
\end{align}
For the case of linearly polarized light $\beta=\gamma$ and the ab-initio Hamiltonian for electrons $H=p^2/2m + V(r)$ where $(\partial_k v)_{ab} =0$ holds for $a \neq b$, 
the coefficient $C_{\alpha\beta\gamma}(\omega)$ for the nonlinear conductivity Eq.\eqref{eq: sigma Im D} is given by
\begin{align}
    C_{\alpha\beta\beta}(\omega)
    &=4\omega
    \left\{\int[dk]  \sum_{ab} 
\t{Im}[g_{ab} v^\beta_{ba}] \frac{f_{ab}}{\epsilon_{ba}^2} \right\} \n
&\times 
\left\{\int[dk]  \sum_{abc} \t{Re}[v^\alpha_{ac} v^\beta_{cb} g_{ba}] \frac{1}{\epsilon_{ac}}\left(\frac{f_{ab}}{\epsilon_{ba}^3} - \frac{f_{cb}}{\epsilon_{bc}^3}\right) \right\},
\end{align}
in the least order in $\omega$.
In general, the coefficient $C_{\alpha\beta\beta}(\omega)$ is proportional to $\omega$ in the presence of $\mathcal{T}$.

\section{Phonon shift current \label{sec: phonon shift current}}
Let us consider the incoming photon frequency is resonant to the phonon frequency, $\omega \simeq \omega_0$, to study phonon shift current.
In this case, we replace $D(\omega)$ by the delta function as
\begin{align}
D(\omega) \simeq 
-i\pi [\delta(\omega-\omega_0)-\delta(\omega+\omega_0)].
\end{align}
Using Eq.~\eqref{eq: sigma Im D} for the nonlinear conductivity, we obtain the formula for the phonon shift current as
\begin{align}
\sigma_{\alpha \beta \gamma}(\omega) 
&\simeq
\pi e^3 \t{Im}\left[ \tilde A^\beta_1(\omega) \tilde A^{\alpha\gamma}_2(\omega)
-
\tilde A^\gamma_1(-\omega) \tilde A^{\alpha\beta}_2(-\omega)
\right] 
\n
& 
\times \delta(\omega-\omega_0),
\end{align}
with $\tilde A_1$ and $\tilde A_2$ in Eq.~\eqref{eq: A1} and Eq.~\eqref{eq: A2}.

For the case of linearly polarized light and the ab-initio Hamiltonian, 
the expression for the phonon shift current is given by
\begin{align}
\sigma_{\alpha \beta \beta}(\omega) &=
4 \pi e^3 \omega
\left\{\int[dk]  \sum_{ab} 
\t{Im}[g_{ab} v^\beta_{ba}] \frac{f_{ab}}{\epsilon_{ba}^2} \right\} \n
&\times 
\left\{\int[dk]  \sum_{abc} \t{Re}[v^\alpha_{ac} v^\beta_{cb} g_{ba}] \frac{1}{\epsilon_{ac}}\left(\frac{f_{ab}}{\epsilon_{ba}^3} - \frac{f_{cb}}{\epsilon_{bc}^3}\right) \right\}
\n
&\times 
\delta(\omega-\omega_0).
\label{eq: sigma}
\end{align}
This formula clearly indicates that  photoexcitations of phonons, which are charge neutral and usually lie much below the electronic band gap, can produce dc charge current generation.
Such charge current is essentially induced by the shift current mechanism since the phonons in noncentrosymmetric crystals accompany nonzero electric polarization.

\section{Direct current from dielectric loss \label{sec: dielectric loss}}

In this section, we consider DC generation arising from dielectric loss in ferroelectrics with off resonant driving in the low frequency region ($\omega \ll \omega_0$).

We start with the expression for the DC generation in Eq.~\eqref{eq: sigma Im D}.
The presence of the imaginary part of the phonon propagator $\t{Im}[D(\omega)]$ in Eq.~\eqref{eq: sigma Im D} describes the energy dissipation due to the phonons, which corresponds to the dielectric loss in the low frequency region.
Thus the formula Eq.~\eqref{eq: sigma Im D} indicates that the dielectric loss can generally induce the DC generation through shift current mechanism.
The imaginary part of the phonon propagator reads
\begin{align}
    \t{Im}[D(\omega)]
    &=\frac{-4\omega_0 \gamma \omega}{(\omega^2 - \omega_0^2)^2+(2\gamma \omega)^2} \n
    &\simeq -4 \frac{\gamma \omega}{\omega_0^3}, \qquad (\omega \ll \omega_0)
\end{align}
in the low frequency region.
Thus the nonlinear conductivity in the low frequency region is given by
\begin{align}
    \sigma_{\alpha\beta\gamma}(\omega) = \frac{4e^3\gamma\omega}{\omega_0^3} C_{\alpha\beta\gamma}(\omega),
\end{align}
with the coefficient $C_{\alpha\beta\gamma}(\omega)$ in Eq. \eqref{eq: C}.
For the time reversal symmetric systems under linearly polarized light $\beta=\gamma$,
the nonlinear conductivity scales as  
\begin{align}
    \sigma_{\alpha\beta\beta}(\omega) \propto \omega^2, \qquad (\omega \ll \omega_0)
\end{align}
because of $C_{\alpha\beta\beta}(\omega) \propto \omega$.
This behavior is reasonable in that the the current should vanish in an insulator with the adiabatic limit $(\omega \to 0)$,
since the electric polarization adiabatically follows the external electric field for $\omega \to 0$. 
In addition, one can interchange $\omega \leftrightarrow -\omega$ in Eq. ~\eqref{eq: J shift} for linearly polarized light $\beta=\gamma$, indicating that $\sigma_{\alpha \beta\beta}(\omega)$ is an even function of $\omega$.
Combining these two properties leads to the scaling $\sigma_{\alpha\beta\beta}(\omega) \propto \omega^2$ in the low frequency region.
In contrast,  for ciruclarly polarized light ($\beta \neq \gamma$), $\omega \leftrightarrow -\omega$ is not interchangable in Eq. ~\eqref{eq: J shift},
indicating the $\omega$ linear term in $\sigma_{\alpha\beta\gamma}(\omega)$ is allowed.

In comparison, at the phonon resonance $\omega=\omega_0$, the DC generation under the linearly polarized light is given by
\begin{align}
    \sigma_{\alpha\beta\beta}(\omega_0) &= \frac{e^3}{\gamma} C_{\alpha\beta\beta}(\omega_0)
    ,
\end{align}
with $\t{Im}[D(\omega_0)]=1/\gamma.$
Thus the ratio of the DC generation associated with dielectric loss to the phonon shift current is given by
\begin{align}
    \frac{\sigma_{\alpha\beta\beta}(\omega)}{\sigma_{\alpha\beta\beta}(\omega_0)}
    &\simeq \frac{4\gamma^2\omega}{\omega_0^3} \frac{C_{\alpha\beta\beta}(\omega)}{C_{\alpha\beta\beta}(\omega_0)}
    =\frac{4\gamma^2\omega^2}{\omega_0^4},
\end{align}
in the low frequency region, $\omega \ll \omega_0$.

To discuss the amount of the DC in the present mechanism incorporating the effect of the absorption depth, it is good to look at the Glass coefficient which is defined by
\begin{align}
    G_{\alpha\beta\beta}(\omega) &= \frac{J^\alpha_\t{dc}}{\alpha_\t{abs} I_\beta},
\end{align}
with the absorption coefficient $\alpha_\t{abs}$ and the intensity of the incident light $I_\beta$ with linear polarization along the $\beta$ direction \cite{Glass74,Osterhoudt19}.
The absorption coefficient $\alpha_\t{abs}$ is proportional to the linear conductivity $\sigma_{\beta\beta}(\omega)$,
which scales linearly in $\omega$, if we neglect relaxation mechanisms other than the phonon excitations.
This $\omega$ linear scaling of $\sigma_{\beta\beta}(\omega)$ can be derived from 
\begin{align}
    \sigma_{\beta\beta}(\omega) 
    &=
    -e^2
    \left\{\tilde A^\beta_1(\omega) \t{Im}[D(\omega)] [A^\beta_{1b}(\omega)]^* +
(\omega \leftrightarrow -\omega)
\right\},
\end{align}
where $\tilde A^\beta_1(\omega) [A^\beta_{1b}(\omega)]^* \to const.$ and $\t{Im}[D(\omega)] \propto \omega$ for $\omega \ll \omega_0$.
Thus the Glass coefficient behaves as
\begin{align}
    G(\omega) \propto \frac{\sigma_{\alpha\beta\beta}(\omega)}{\sigma_{\beta\beta}(\omega)} 
    \propto \omega,
\end{align}
in the low frequency region.

\section{Application to Rice-Mele model \label{sec: Rice Mele}}

\begin{figure}[t]
    \centering
    \includegraphics[width = \linewidth]{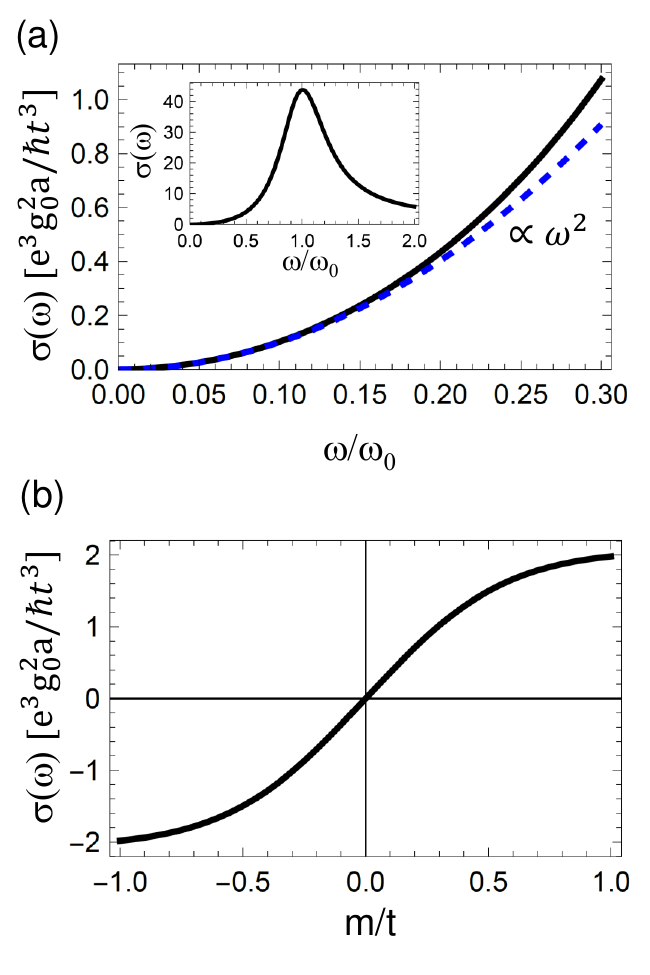}
    \caption{
    DC generation in Rice-Mele model coupled with phonons.
    (a) Nonlinear conductivity $\sigma(\omega)$ in the low frequency regime below the phonon excitation. Blue dashed curve is a fit of $\sigma(\omega)$ with a form $C\omega^2$ indicating the scaling of $\sigma(\omega) \propto \omega^2$.
    Inset shows $\sigma(\omega)$ in the phonon resonance region $\omega \sim \omega_0$ corresponding to the phonon shift current.
    (b) Staggered potential ($m$) dependence of nonlinear conductivity $\sigma(\omega)$. $\sigma(\omega)$ at the fixed frequency, $\omega=0.25\omega_0$, shows a sign change with respect to the sign of $m$, indicating that the current direction switches depending on the direction of the electric polarization. 
    We adopted the parameters, $\delta t/t=0.2, \omega_0/\omega=0.2, \gamma/t=0.05$ for (a) and (b), and $m/t=0.2$ for (a).
    }
    \label{fig: numerics}
\end{figure}

In this section, we demonstrate the DC generation by dielectric loss by applying our formulation to a model of ferroelectrics.
Specifically, we consider Rice-Mele model which is a representative model of 1D ferroelectrics and is described by the Hamiltonian given by
\begin{align}
H_{RM}=
\sum_k
\Psi^\dagger_{k}
\left(t \cos \frac{k a}{2} \sigma_x + \delta t \sin \frac{k a}{2} \sigma_y + m \sigma_z \right)
\Psi_{k},
\label{eq: RM}
\end{align}
with 
\begin{align}
    \Psi_{k}
    =
    \begin{pmatrix}
    c_{A,k} \\ c_{B,k}
    \end{pmatrix},
\end{align}
where $t$ is the hopping amplitude, $\delta t$ is the strength of hopping alternation, $m$ is the strength of staggered potential, $a$ is the lattice constant, and $\sigma_i$ is the Pauli matrix acting on sublattice (AB) degrees of freedom.
We consider an electron-phonon coupling in the form of Eq.~\eqref{eq: Helph q=0} with
\begin{align}
g(k)&= g_0 \sin \frac{k a}{2} \sigma_y,
\end{align}
where the phonon excitation modulates the bond alternation with the strength $g_0$.

In this setup, the diagrams constituting DC generation in Fig.~\ref{fig: diagrams} can be evaluated with Eq.~\eqref{eq: A1} and Eq.~\eqref{eq: A2}.
We show the reuslt of numerical calculation in Fig.~\ref{fig: numerics} for the parameters.
Figure \ref{fig: numerics}(a) shows the nonlinear conductivity $\sigma(\omega)$ in the low frequency regime below the phonon excitation, where DC generation by the dielectric loss takes place. 
In the low frequency region, $\sigma(\omega)$ scales as $\propto \omega^2$ as indicated by a $\omega^2$ fit in the blue dashed curve.
The inset shows a behavior of $\sigma(\omega)$ in the phonon resonance region $\omega \sim \omega_0$ which corresponds to the phonon shift current.
The nonlinear conductivity $\sigma(\omega)$ at fixed $\omega$ shows a sign change with respect to the sign of the staggered potential $m$ as shown in Fig.~\ref{fig: numerics}(b).
Since $m$ determines the pattern of the inversion symmetry breaking and the direction of the electric polarization, this sign change of $\sigma(\omega)$ indicates that the direction of the DC generation is governed by the direction of the electric polarization.

\section{Discussion \label{sec: discussion}}

Berry phase was originally formulated for the adiabatic 
process, where the transitions of wavefunction between the different
energy eigenstates are forbidden \cite{Berry1990}.
Therefore, its 
application to solids is naively considered to be limited to the 
low energy phenomena where the inter-band transitions are forbidden,
i.e., the wavefunctions are confined within the manifold spanned by 
the states specified by the band index \cite{Xiao10}.
On the other hand, the shift current due to
the interband transition is formulated by the Berry connections of 
Bloch wavefunctions, where the adiabatic approximation is never justified.
In the case of the shift current due to phonon excitation,
there is no real transitions of electrons between bands; instead
the virtual transition of electrons that accompanies the real excitation of phonons 
induces the DC.  
Thus, it appears to be an ``adiabatic'' process for 
electronic degrees of freedom. However, from the viewpoint of the Born-Oppenheimer
approximation (BOA) \cite{Mead92}, which is the adiabatic approximation based on 
the difference in the time-scales of electrons and phonons,
the nonadiabatic corrections to BOA is the origin of the
shift current. 
This is especially clear in the present 
dielectric loss, in that the small but finite frequency 
gives the absorption and shift current simultaneously.
Also, the existence of energy supply
is crucial for shift current responses to support the DC generation in non-superconducting system
even though it is of a geometric origin.

Note that, in real materials, the dielectric loss could occur also due to the
motion of the domain walls. In this case, the generation of 
the shift current is localized in the domain wall regions, and 
has the opposite signs between the two types of the
domain wall, i.e., $(+,-)$ and $(-,+)$.

Lastly, we present an estimation for the DC generation from dielectric loss.
For soft phonon excitations in BaTiO$_3$ studied in Ref. \cite{Okamura22},
photocurrent of the order of 10~$\mu$A with the Glass coefficient $G \simeq 1 \times 10^{-8} $~cm/V has been observed.
For soft phonons (e.g., Slater mode in BaTiO$_3$), the typical values for the phonon frequency and the energy broadening are given by $\omega_0 \simeq 4$~meV and $\gamma \simeq 4$~meV \cite{Okamura22}.
If one considers the dielectric loss in the presence of AC electric 
field in the off resonant gigahertz regime, e.g., $\omega=10$~GHz, 
the ratio of the nonlinear conductivities $\sigma_{\alpha\beta\beta}(\omega)/\sigma_{\alpha\beta\beta}(\omega_0)$ is given by $4\gamma^2\omega^2/\omega_0^4 \simeq 4 \times 10^{-4}$.
Since the nonlinear conductivity of $\sigma_{\alpha\beta\beta}(\omega_0) \simeq 30~\si{\mu A/V^2}$ is reported in ab-initio calculation \cite{Okamura22}, we expect $\sigma_{\alpha\beta\beta}(\omega) \simeq 0.01~\si{\mu A/V^2}$ for $\omega =10 ~\si{GHz}$.
For the electric field of $E=1~\si{kV/cm}$ and the sample dimension of $d=0.1~\si{mm}$, the DC generation of $1~\si{\mu A}$ is expected from dielectric loss, which is feasible for experimental detection.
For the Glass coefficient, we expect $G \simeq 1 \times 10^{-10}~\si{cm/V}$ for $\omega =10~\si{GHz}$ from the measured value for the phonon shift current,  $G\simeq 1 \times 10^{-8}~\si{cm/V}$ for $\omega =1 ~\si{THz}$.

\acknowledgements
We thank N. Ogawa, K. Kanoda and Y. Tokura for fruitful discussions.
This work was supported by 
JSPS KAKENHI Grant 23H01119, 23K17665 (T.M.), and
JST CREST (Grant No. JPMJCR19T3) (T.M.).

\bibliography{ref}

\end{document}